\documentclass[nopacs,epj]{svjour}
\usepackage{graphicx}

\usepackage{epstopdf}

\usepackage{amssymb}
\usepackage{amsmath}

\usepackage{multirow}

\usepackage{color}

\usepackage[colorlinks]{hyperref}

\AtBeginDocument{%
    \hypersetup{
      urlcolor     = blue, 
      linkcolor    = black, 
      citecolor   = blue, 
      pdfborder={0 0 0},
      urlbordercolor={1 1 1},
      citebordercolor={1 1 1}
    }
    } 

\begin{document}

\hyphenation{coun-ter cor-res-pon-din-gly e-xam-ple
co-o-pe-ra-tion ex-pe-ri-men-tal pum-ping ve-ri-fi-ca-tion ve-ri-fi-ca-tions ha-ving pro-ba-bi-li-ty A-lice}

\title{A novel delayed-choice experimental proposal testing local decisions}


\author{Stefan~Ataman\inst{1}
}                     

\institute{Extreme Light Infrastructure - Nuclear Physics
(ELI-NP), 30 Reactorului Street, 077125 Magurele, 
Romania \email{stefan.ataman@eli-np.ro} }

\date{Received: date / Revised version: date}
%
\abstract{Entangled states are notoriously non-separable, their sub-ensembles being only statistical mixtures yielding no coherences and no quantum interference phenomena. The interesting features of entangled states can be revealed only by coincidence counts over the (typically) two sub-ensembles of the system. In this paper we show that this feature extends to properties thought to be local, for example the transmissivity coefficient of a beam splitter. We discuss a well-known experimental setup and propose modifications, so that delayed-choice can be added and this new feature of entanglement tested.
%
\PACS{
      {PACS-key}{describing text of that key}   \and
      {PACS-key}{describing text of that key}
     } 
} 
%


\maketitle
%

\section{Introduction}
\label{sec:introduction}
Entanglement is probably the most counter-intuitive yet the most important distinctive feature of the Quantum World. Ignited by the famous Einstein-Podolsky-Rosen paper \cite{Ein35}, brought from philosophical debates into the lab by John Bell \cite{Bel64}, entanglement is nowadays a practical resource \cite{Bou00}.


The coherent superposition of a single photon Fock state impinging at both inputs of on a beam splitter is well known and well studied \cite{Yur86,Ou87,Lou03}. This is done typically with a Mach-Zehnder interferometer and the detection rates of the single-photon counters depends sinusoidally on the path length difference between the two arms of the interferometer. However, if the beam splitter is not balanced, the visibility of the interference fringes diminishes. Therefore, if one computes the ratio of the detection rates of the two output detectors, it heavily depends on both the path length difference of the two arms and the transmissitivity of the output beam splitter. This can be explained through Bohr's complementarity principle \cite{Boh28} and inequalities quantifying wave-like and particle-like behavior have been derived \cite{Woo79,Dur00,Gre88,Eng95,Eng96}. 


The most popular source of single photons today is the process of spontaneous parametric down-conversion (SPDC) \cite{Bur70}. In the 1980's and 1990's, Mandel's group proposed and performed ground-breaking experiments involving two non-linear crystals. In such an experiment\cite{Ou90a,Ou90b}, ``phase memory'' of the pump beam implied an ``induced coherence'' between the two crystals: although the process of SPDC is random, coherence can be induced \cite{Men16}. A follow-up proposal by Ou \cite{Ou97} showed how this type of experiment can be extended into a quantum eraser \cite{Scu82,Scu91}.

The same group reported a special kind of experiment \cite{Zou91}. This time, the idler beams of the two non-linear crystals overlap, therefore a detection at the idler photo-detector could not determine which crystal emitted the photon pair. Dubbed ``mind-boggling'', this experiment displayed (or not) interference between signal beams in function of the absence (or presence) of a object blocking the idler beams. More than twenty years later, this principle launched a new research field: quantum imaging \cite{Lem14,Kal16,Ata16a,Pat16}.

The idea of delayed-choice originated with Wheeler's seminal papers \cite{Whe78,Whe83} and was confirmed only decades later in an experiment \cite{Jac07} involving fully space-like separated decisions based on a quantum random number generator. Quantum elaborations of the initial experiment have been proposed \cite{Ion11} and experimentally confirmed \cite{Kai12}. In a nutshell, delayed-choice experiments prove that there is no decision (``wave'', ``particle'' or partially both) taken by a particle before measurement.

In this paper we will use the idea of delayed-choice for a different purpose: we show that even local properties like the transmissivity of a beam splitter can become -- in a way -- non-local. Moreover, delayed-choice is added, therefore \emph{information} about this property should not be available at arbitrary distances. Yet, the only way to reveal it is through coincidence counts, even if information about the choice taken could not have propagated from one part of the experiment to the other.

This paper is structured as follows. In Section \ref{sec:interference_var_trans_BS} we consider a coherent superposition of single photon Fock states impinging on a variable transmissitivity beam splitter. Entangled states are considered in Section \ref{sec:interference_engantled_state}, where the effect of entanglement is discussed for both local (single) and non-local (coincidence) measurements. An experimental setup displaying the theoretically discussed effect is described in Section \ref{sec:Experimental_setups}. The delayed-choice version of the same experiment is proposed in Section \ref{sec:OWZM_delayed_choice} and finally conclusions are drawn in Section \ref{sec:conclusions}.

\section{A coherent superposition incident on beam splitter}
\label{sec:interference_var_trans_BS} 

\begin{figure}
\centering
\includegraphics[width=2.5in]{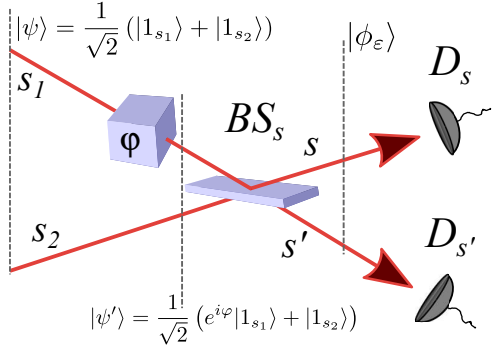}
\caption{A coherent superposition of single photon states is incident on the beam splitter $\text{BS}_s$ having a variable transmissivity $T=\varepsilon$. The output detection ratio of the two photo-counters ($D_s$ and $D_{s'}$) heavily depends on $\varepsilon$ and $\varphi$.} \label{fig:Coherent_superposition_beam_splitter}
\end{figure}

\begin{figure}
	\centering
	\includegraphics[width=3.2in]{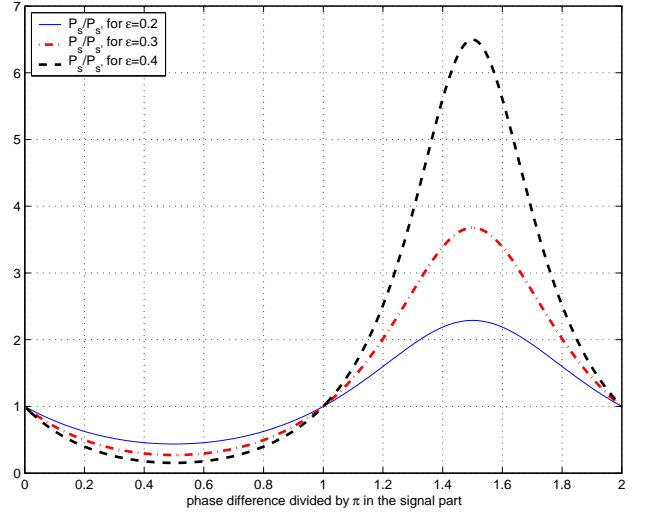}
	\caption{The ratio ${P_{s}}/{P_{s'}}$ from equation \eqref{eq:P_detection_ratio_s_s_prime} for three values of $\varepsilon$. As the beam splitter transmissivity coefficient $\varepsilon\to1/\sqrt{2}$, the ratio  ${P_{s}}/{P_{s'}}$ shows higher and higher amplitude variations.}
	\label{fig:Ratio_Ds_D_sprime_octave}
\end{figure}

Suppose we have a state vector describing a coherent superposition of two single-photon Fock states in modes $s_1$ and, respectively, $s_2$,
\begin{equation}
\label{eq:cho_superposition_initial}
\vert\psi\rangle=\frac{1}{\sqrt{2}}\left(\vert1_{s_1}\rangle+\vert1_{s_2}\rangle\right)
\end{equation}
We call this the ``signal'' wavevector. Throughout this paper, all modes are assumed to be monochromatic, all beam splitters lossless and all photo-detectors ideal. Before hitting the beam splitter $\text{BS}_s$ (see Fig.~\ref{fig:Coherent_superposition_beam_splitter}) the quantum state $\vert\psi\rangle$ evolves into
\begin{equation}
\label{eq:cho_superposition_evolved}
\vert\psi'\rangle=\frac{1}{\sqrt{2}}\left(e^{i\varphi}\vert1_{s_1}\rangle+\vert1_{s_2}\rangle\right)
\end{equation}
due to a voluntarily introduced delay (modeled through the phase $\varphi$) in the first path. We characterize the beam splitter $\text{BS}_s$ by the transmissivity (reflectivity) coefficient $T=\varepsilon$ ($R=i\sqrt{1-\varepsilon^2}$) with $\varepsilon\in[0,1]$. The quantum state after the beam splitter is easily computed yielding
\begin{eqnarray}
\label{eq:phi_simple_quantum_sys_varepsilon}
\vert\phi_\varepsilon\rangle=\frac{\varepsilon+ie^{i\varphi}\sqrt{1-\varepsilon^2}}{\sqrt{2}}\vert1_s\rangle
+\frac{i\sqrt{1-\varepsilon^2}+\varepsilon e^{i\varphi}}{\sqrt{2}}\vert1_{s'}\rangle
\end{eqnarray}
The probability of photo-detection at the detectors $D_s/D_{s'}$ is found to be
\begin{equation}
\label{eq:P_detection_s_s_prime_varepsilon}
P_{s/s'}=\frac{1\mp2\varepsilon\sqrt{1-\varepsilon^2}\sin\varphi}{2}=\frac{1\mp{V_\varepsilon}\sin\varphi}{2}
\end{equation}
and it shows the famous interference fringes with a visibility
\begin{equation}
\label{eq:single_system_interference_fringe_visibility}
V_\varepsilon=2\varepsilon\sqrt{1-\varepsilon^2}
\end{equation}
The ratio of photo-counts at the detectors $D_s$ and $D_{s'}$ is given by
\begin{equation}
\label{eq:P_detection_ratio_s_s_prime}
\frac{P_{s}}{P_{s'}}=\frac{1-{V_\varepsilon}\sin\varphi}{1+{V_\varepsilon}\sin\varphi}
\end{equation}
and, as seen in Fig.~\ref{fig:Ratio_Ds_D_sprime_octave}, it heavily depends on the beam splitter's transmissivity coefficient ($\varepsilon$) and on the path length difference, modelled here by the phase difference $\varphi$. The maximum for the photo-detection rate ratio \eqref{eq:P_detection_ratio_s_s_prime} is obtained for $\sin\varphi=1$ and it raises at high values as the beam splitter becomes balanced (i.e. as $\varepsilon\to1/\sqrt{2}$). This is the behavior one expects from the coherent superposition of single-photon states \eqref{eq:cho_superposition_initial} applied to a beam splitter having a certain transmissitivity coefficient, $\varepsilon$.

\section{Single and coincidence detection rates for an entangled state}
\label{sec:interference_engantled_state}
We now assume that the quantum state from equation \eqref{eq:cho_superposition_initial} became entangled with a second system with wavevectors denoted by $\vert1_{i_1}\rangle$ and $\vert1_{i_2}\rangle$ (the ``idler'' part). If the state is maximally entangled, it can be written as
\begin{equation}
\label{eq:Psi_s1_i1_s2_i2_initial}
\vert\Psi\rangle=\frac{1}{\sqrt{2}}\left(\vert1_{s_1}\rangle\otimes\vert1_{i_1}\rangle+\vert1_{s_2}\rangle\otimes\vert1_{i_2}\rangle\right)
\end{equation}
and if we assume the same evolution of the signal part as described in Section \ref{sec:interference_var_trans_BS}, we have 
\begin{equation}
\label{eq:Psi_s1_i1_s2_i2_evolved}
\vert\Psi'\rangle=\frac{1}{\sqrt{2}}\left(e^{i\varphi}\vert1_{s_1}\rangle\otimes\vert1_{i_1}\rangle+\vert1_{s_2}\rangle\otimes\vert1_{i_2}\rangle\right)
\end{equation}
After the beam splitter $\text{BS}_s$ the quantum state evolves to
\begin{eqnarray}
\label{eq:Phi_s1_i1_s2_i2_only_s}
\vert\Phi_\varepsilon\rangle=\frac{i{e^{i\varphi}}\sqrt{1-\varepsilon^2}}{\sqrt{2}}\vert1_s\rangle\otimes\vert1_{i_1}\rangle
+\frac{\varepsilon{e^{i\varphi}}}{\sqrt{2}}\vert1_{s'}\rangle\otimes\vert1_{i_1}\rangle
\nonumber\\
+\frac{\varepsilon}{\sqrt{2}}\vert1_{s}\rangle\otimes\vert1_{i_2}\rangle
+\frac{i\sqrt{1-\varepsilon^2}}{\sqrt{2}}\vert1_{s'}\rangle\otimes\vert1_{i_2}\rangle
\end{eqnarray}
If one wants to compute the single detection rates $P_s$ and $P_{s'}$ again, then a partial trace over the ``idler'' part of the wavevector $\vert\Phi_\varepsilon\rangle$ has to be done. We first build the density matrix $\hat{\rho}=\vert\Phi_{\varepsilon}\rangle\langle\Phi_{\varepsilon}\vert$ of the global system, then partially trace over the idler modes yielding
\begin{eqnarray}
\label{eq:rho_signal_max_entangled}
\hat{\rho}_s
=\text{Tr}_{\{i_1,i_2\}}\left\{\hat{\rho}\right\}
=\frac{1}{2}\vert1_s\rangle\langle1_s\vert
+\frac{1}{2}\vert1_{s'}\rangle\langle1_{s'}\vert
\end{eqnarray}
and the probabilities of photo-detection at the detectors  $D_s$ and $D_{s'}$ are straightforward to compute,
\begin{eqnarray}
\label{eq:Prob_phi1_rho_s_Q_eraser_perfect_marker}
P_{s}=\langle{1_s}\vert\hat{\rho}_s\vert1_s\rangle=\frac{1}{2}
=\langle{1_{s'}}\vert\hat{\rho}_s\vert1_{s'}\rangle=P_{s'}
\end{eqnarray}
therefore
\begin{equation}
\label{eq:P_detection_s_s_prime_is_1}
\frac{P_{s}}{P_{s'}}=1
\end{equation}
whatever the value of the transmissivity of the beam splitter $\text{BS}_s$ is, in stark contrast with what we found in equation \eqref{eq:P_detection_ratio_s_s_prime}. It is noteworthy that the ratio from equation \eqref{eq:P_detection_s_s_prime_is_1} doesn't depend on the path length difference $\varphi$, either.

If one assumes now Alice (Bob) operating the signal (idler) part and of the wavevector \eqref{eq:Psi_s1_i1_s2_i2_initial}, they have good reasons to be surprised: Alice expects to have a trace in her measurements  of the transmissivity coefficient ($\varepsilon$) she has chosen for the beam splitter $\text{BS}_s$. Yet, as it can be seen from the photo-counts she registers at $D_s$ and $D_{s'}$, there is no trace of it. 

Only correlations with Bob (who is totally unaware of Alice's decisions on $\varepsilon$) can reveal Alice's local choice. From equation \eqref{eq:Phi_s1_i1_s2_i2_only_s} one can immediately compute the coincidence photo-detection probabilities at the detectors $D_s-D_{i_1}$ and $D_{s'}-D_{i_1}$ yielding
$P_{s,i_1}=(1-\varepsilon^2)/2$ and, respectively, $P_{s',i_1}=\varepsilon^2/2$. The ratio of the coincidence counts
\begin{equation}
\label{eq:P_coincidence_ratio_s_i1_sprime_i1}
\frac{P_{s,i_1}}{P_{s',i_1}}=\frac{1-\varepsilon^2}{\varepsilon^2}
\end{equation}
depends on the transmissitivity ratio $\varepsilon$, although in a different manner compared to equation $\eqref{eq:P_detection_s_s_prime_varepsilon}$. However, no coincidence (or single) counts can depend on $\varphi$ if one considers the state vector \eqref{eq:Phi_s1_i1_s2_i2_only_s}. This is so because $i_1$ and $i_2$ are perfect which-path markers and $\varphi$ is connected to the wave-like character of the wavefunction. One could recover also $\varphi$ if we ``erase'' the which-path information. This can be done by a second beam splitter, $\text{BS}_i$ having the transmissivity (reflectivity) coefficient $T=\chi$ ($R=i\sqrt{1-\chi^2}$) with $\chi\in[0,1]$. Denoting the output modes of the beam splitter $\text{BS}_i$ by $i$ and $i'$ ($i_1$ goes through transmission to $i'$) one finds the state vector
\begin{eqnarray}
\label{eq:Phi_s_sprime_i_iprime}
\vert\Phi_{\varepsilon,\chi}\rangle
=\frac{-e^{i\varphi}\sqrt{1-\varepsilon^2}\sqrt{1-\chi}+\varepsilon\chi}{\sqrt{2}}
\vert1_s\rangle\otimes\vert1_{i}\rangle
\nonumber\\
+i\frac{e^{i\varphi}\varepsilon\sqrt{1-\chi^2}+\sqrt{1-\varepsilon^2}\chi}{\sqrt{2}}\vert1_{s'}\rangle\otimes\vert1_{i}\rangle
\nonumber\\
+i\frac{e^{i\varphi}\sqrt{1-\varepsilon^2}\chi+\varepsilon\sqrt{1-\chi^2}}{\sqrt{2}}\vert1_{s}\rangle\otimes\vert1_{i'}\rangle
\nonumber\\
+\frac{e^{i\varphi}\varepsilon\chi-\sqrt{1-\varepsilon^2}\sqrt{1-\chi^2}}{\sqrt{2}}
\vert1_{s'}\rangle\otimes\vert1_{i'}\rangle
\end{eqnarray}
The coincidence rates at the detectors $D_s-D_{i}$ and $D_{s'}-D_{i}$ are found to be
\begin{equation}
\label{eq:P_coincidence_s_i_eps_chi}
P_{s,i}=\frac{1-\varepsilon^2-\chi^2+2\varepsilon^2\chi^2-\chi\sqrt{1-\chi^2}V_\varepsilon\cos\varphi}{2}
\end{equation}
and
\begin{equation}
\label{eq:P_coincidence_sprime_i_eps_chi}
P_{s',i}=\frac{\varepsilon^2+\chi^2-2\varepsilon^2\chi^2+\chi\sqrt{1-\chi^2}V_\varepsilon\cos\varphi}{2}
\end{equation}
where $V_\varepsilon$ is given by equation \eqref{eq:single_system_interference_fringe_visibility}. The ratio of the coincidence counts yields
\begin{equation}
\label{eq:P_coincidence_ratio_s_i_sprime_i}
\frac{P_{s,i}}{P_{s',i}}=\frac{1-\varepsilon^2-\chi^2+2\varepsilon^2\chi^2-\chi\sqrt{1-\chi^2}V_\varepsilon\cos\varphi}{\varepsilon^2+\chi^2-2\varepsilon^2\chi^2+\chi\sqrt{1-\chi^2}V_\varepsilon\cos\varphi}
\end{equation}
For $\chi=0$ one immediately finds the result from equation \eqref{eq:P_coincidence_ratio_s_i1_sprime_i1}. Once again, a coincidence count between Alice and Bob can reveal the parameter $\varepsilon$ (but not $\varphi$). If $\chi=1/\sqrt{2}$ (i. e. $\text{BS}_i$ is balanced), we find the result from equation \eqref{eq:P_detection_ratio_s_s_prime} (with the replacement $\varphi\to\varphi+\pi/2$) and both $\varepsilon$ and $\varphi$ can be observed by Alice. This time, Bob's beam splitter acted as a perfect quantum eraser \cite{Scu82,Scu91} therefore no which-path information was available \cite{Ata14b,Ata16b}.

The new and surprising fact here is the impossibility of Alice to find traces of her own experimental parameter -- the transmissivity $\varepsilon$ of the beam splitter $\text{BS}_s$ -- in any single detection rate. She can find this parameter either if the initial state is unentangled (see Appendix \ref{sec:app:A}) or, if she considers coincidence measurements with Bob. The puzzling feature is that local parameters that should be connected with Alice only reveal themselves uniquely in coincidence measurement. This feature will be taken one step further in Section \ref{sec:OWZM_delayed_choice}, where delayed choice is added in choosing the parameter $\varepsilon$.

\begin{figure}
	\includegraphics[scale=0.45]{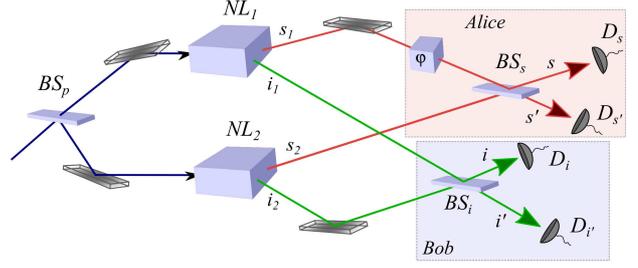}
	\caption{The experimental setup for a maximally entangled state $\vert\Psi\rangle$ given by equation $\eqref{eq:Psi_s1_i1_s2_i2_initial}$. The beam splitter $\text{BS}_s$ is characterized by the transmissivity (reflectivity) coefficient $T=\varepsilon$ ($R=i\sqrt{1-\varepsilon^2}$) while the beam splitter $\text{BS}_i$ is characterized by the transmissivity (reflectivity) coefficient $T=\chi$ ($R=i\sqrt{1-\chi^2}$). Alice (Bob) controls the beam splitter $\text{BS}_s$ ($\text{BS}_i$) and reads the detectors $D_s/D_{s'}$ ($D_i/D_{i'}$).}
	\label{fig:OWZM_type_experiment}
\end{figure}

\section{Experimental setup with two non-linear crystals}
\label{sec:Experimental_setups}
The maximally entangled state from  equation \eqref{eq:Psi_s1_i1_s2_i2_initial} can be obtained with the experimental setup from Fig.~\ref{fig:OWZM_type_experiment}. It is the experimenal setup propozed by Ou, Wang, Zhou and Mandel (OWZM) \cite{Ou90a,Ou90b}. It is noteworthy that although in this experimental setup a single pair of photons is created, this pair is in a coherent superposition of originating from the first \emph{and} second non-linear crystal. Coherence is induced by the pumping laser \cite{Ou90a,Ou90b,Men16}.

As discussed in the previous section, due to signal-idler entanglement, no variation of the single detection rate ratio $P_s/P_{s'}$ should be observed on varying $\varphi$ (with the idler detectors ignored). This should happen regardless of the value of $\varepsilon$.

However, by using coincidence counts, a whole range of ratios $P_{s,i}/P_{s',i}$ can be obtained. As predicted by equation \eqref{eq:P_coincidence_ratio_s_i_sprime_i}, by varying Bob's beam splitter parameter $\chi$, one can continuously morph between ``wave-like'' and ``particle-like'' behavior of our system. 
Equation \eqref{eq:P_coincidence_ratio_s_i_sprime_i} is satisfied whatever the values of $\varepsilon$ and $\chi$, while equations \eqref{eq:P_detection_ratio_s_s_prime} and, respectively \eqref{eq:P_coincidence_ratio_s_i1_sprime_i1} are two extremes (total wave-like, and, respectively, total particle-like behavior).

\section{A delayed-choice experimental proposal}
\label{sec:OWZM_delayed_choice}

Considering again the OWZM-type experimental setup (see Fig.~\ref{fig:OWZM_type_experiment}), one can add Wheeler-style delayed choice \cite{Whe78,Whe83}, so that Alice chooses $\varepsilon$ \emph{after} Bob has already detected the idler photon. This time \emph{information} about what parameter $\varepsilon$ Alice has chosen couldn't have reached Bob before the detection of the idler photon.

Nonetheless, Alice sees no trace of $\varepsilon$ in any single detection rate, as discussed in Section \ref{sec:interference_engantled_state}. Only by correlating her results with Bob's, she can reveal the beam splitter's parameter, $\varepsilon$. But this information \emph{should be local} to Alice only, having no time to propagate. Yet, distant correlations reveal it while local detections return blank results.

Therefore, we propose a modification of the OWZM-type experiment from Fig.~\ref{fig:OWZM_type_experiment} so that the idler paths are much shorter than the signal ones. This time, Bob's detections (at $D_i/D_{i'}$) will be much in advance compared to the signal part events, operated by Alice.

\begin{figure}
	\includegraphics[scale=0.5]{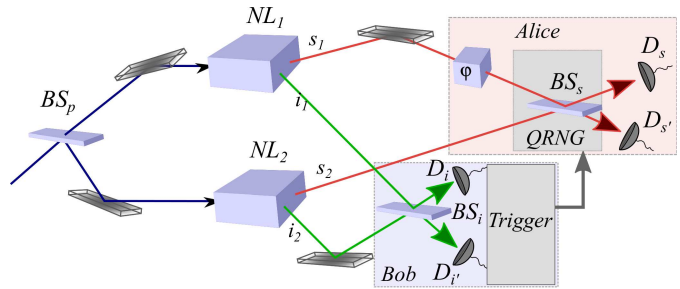}
	\caption{The proposed OWZM-type delayed-choice  experimental setup. Bob's idler part is much shorter than Alice's signal path, so that Bob's detections can trigger early enough a quantum random number generator (QRNG) used to choose the $\varepsilon$ parameter of $\text{BS}_s$. Alice's photon can thus arrive at $\text{BS}_s$ \emph{after} this choice has been done.}
	\label{fig:OWZM_delayed_choice_experiment}
\end{figure}

The experiment is performed as follows: after Bob records an event at either of the detectors, he sends a trigger signal to Alice (see Fig.~\ref{fig:OWZM_delayed_choice_experiment}). Alice receives this signal and starts a quantum random number generator (QRNG) \cite{Ma15}. This will choose $\varepsilon$ for her beam splitter, $\text{BS}_s$. It is supposed that the (signal) optical paths from the non-linear crystals to $\text{BS}_s$ are long enough, so that the decision Alice takes happens before the (signal) photon arrives at the beamsplitter $\text{BS}_s$.

Alice varies $\varphi$ in small steps and for each value of $\varphi$ she records a (large enough) number of detections at $D_s/D_{s'}$. After the experiment is over, Alice can compare the ratio of detections found at the detectors $D_s$ and $D_{s'}$. When considering only her local measurements, she should see a flat result, whatever the values of $\varphi$ and $\varepsilon$.
 However, as discussed in Section \ref{sec:interference_engantled_state}, if Alice correlates her results with Bob's, the coincidence ratios (e. g.  $D_{s,i}/D_{s',i}$ ) should reveal her locally chosen parameter $\varepsilon$.


\section{Conclusions}
\label{sec:conclusions}
In this paper we have shown that if one is bound to entangled states, local properties can become non-local, where only coincidence measurements reveal the given local property (in this case the transmissitivity of a beam splitter).

We proposed an experiment where this property takes the form of an information paradox. The choice of a local parameter - taken as late as possible by Alice -- \emph{cannot} be detected locally. It becomes nonetheless apparent in Alice-Bob correlations, however Bob \emph{cannot} possibly \emph{have any information} about Alice's delayed-manner and locally chosen parameter.

We can conclude that we pushed entanglement induced quantum non-locality one step further in this paper. Although thoroughly studied already, it seems that quantum entanglement has still more stories to tell.

\appendix

\section{A non-entangled state with two crystals}
\label{sec:app:A}
Of course, Alice could recover the results from Section \ref{sec:interference_var_trans_BS} if the initial state \eqref{eq:Psi_s1_i1_s2_i2_initial} is not entangled i. e.
\begin{equation}
\label{eq:Psi_s1_s2_i_NONENTANGLED_initial}
\vert\Psi_n\rangle=\frac{1}{\sqrt{2}}\left(\vert1_{s_1}\rangle+\vert1_{s_2}\rangle\right)\otimes\vert1_{i}\rangle
\end{equation}
This time the same quantum evolution from Section \ref{sec:interference_var_trans_BS} leading to $\vert\Psi'_n\rangle={1}/{\sqrt{2}}\left(e^{i\varphi}\vert1_{s_1}\rangle+\vert1_{s_2}\rangle\right)\otimes\vert1_{i}\rangle$ and $\vert\Phi_{n}\rangle=\vert\psi_\varepsilon\rangle\otimes\vert{i}\rangle$ after the beam splitter $\text{BS}_s$ is to be expected. Therefore, the density matrix of the global system can be written as
\begin{equation}
\hat{\rho}_n=\vert\Phi_{n}\rangle\langle\Phi_{n}\vert=\vert\phi_\varepsilon\rangle\otimes\vert{i}\rangle\langle{i}\vert\otimes\langle\phi_\varepsilon\vert
=\vert\phi_\varepsilon\rangle\langle\phi_\varepsilon\vert\otimes\vert{i}\rangle\langle{i}\vert
\end{equation}
which is clearly separable into signal and idler parts. Therefore, if Alice decides to measure the ratio of single detection rates $D_s/D_{s'}$, she will obtain again the result from equation \eqref{eq:P_detection_ratio_s_s_prime}.

\begin{figure}
	\includegraphics[scale=0.5]{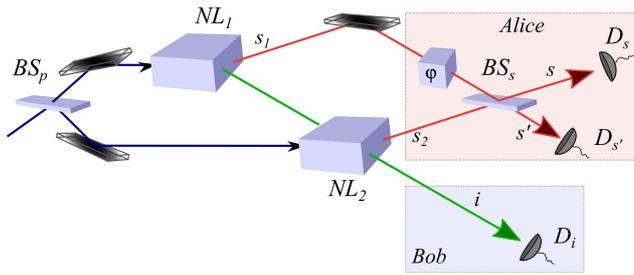}
	\caption{The proposed experimental setup for a non-entangled state $\vert\Psi_n\rangle$ given by equation $\eqref{eq:Psi_s1_s2_i_NONENTANGLED_initial}$. The beam splitter $\text{BS}_s$ is characterized by the transmissivity (reflectivity) coefficient $T=\varepsilon$ ($R=i\sqrt{1-\varepsilon^2}$).}
	\label{fig:ZWM_type_experiment}
\end{figure}

The non-entangled state vector from equation \eqref{eq:Psi_s1_s2_i_NONENTANGLED_initial} can be obtained using the experimental setup depicted in Fig.~\ref{fig:ZWM_type_experiment}. This experiment was proposed by Zhou, Wang and Mandel (ZWM) \cite{Zou91}. This time, if Alice measures the photo-counters $D_s$ and $D_{s'}$, the photo-detection ratio $P_s/P_{s'}$ should depend on the transmissivity $\varepsilon$ of $\text{BS}_s$ and of the phase difference $\varphi$ as predicted by equation \eqref{eq:P_detection_ratio_s_s_prime}. This result is intuitive and both Alice and Bob should not be surprized by their findings. One should not disregard, thought, the fact that everything started with a separable state \eqref{eq:Psi_s1_s2_i_NONENTANGLED_initial}, hence Alice's and Bob's subsystems have separate quantum evolutions.




\begin{thebibliography}{6}

\bibitem{Ein35}
A. Einstein, B. Podolsky, N. Rosen, 
Phys. Rev. \textbf{47}, 777 (1935)

\bibitem{Bel64}
J. S. Bell, 
Physics \textbf{1} 195 (1964)

\bibitem{Bou00}
D. Bouwmeester, A. Ekert, A. Zeilinger, 
\emph{The Physics of Quantum Information: Quantum Cryptography, Quantum Teleportation, Quantum Computation}, 
Springer (2000)

\bibitem{Yur86}
B. Yurke, S. L. McCall, J. R. Klauder, 
Phys. Rev. A \textbf{33}, 4033 (1986)

\bibitem{Ou87}
Z. Y. Ou, C. K. Hong, L. Mandel, 
Opt. Comm. \textbf{63}, 118 (1987)


\bibitem{Lou03}
R. Loudon, \emph{The Quantum Theory of Light}, (Oxford University
Press, Third Edition, 2003)


\bibitem{Boh28}
N. Bohr, 
Naturwissenschaften \textbf{16}, 245 (1928)



\bibitem{Woo79}
W. Wootters, W. H. Zurek, 
Phys. Rev. D \textbf{19}, 473 (1979)

\bibitem{Dur00}
S. D\"urr, G. Rempe, 
Am. J. Phys. \textbf{68} , 1021 (2000)


\bibitem{Gre88}
D. M. Greenberger, A. Yasin, 
Phys. Lett. A \textbf{128}, 391 (1988)


\bibitem{Eng95}
B.-G. Englert, M. O. Scully, H. Walther, 
Nature \textbf{375}, 367 (1995)

\bibitem{Eng96}
B.-G. Englert, 
Phys. Rev. Lett. \textbf{77}, 2154 (1996)


\bibitem{Bur70}
D.C. Burnham, D.L. Weinberg, Phys. Rev. Lett. \textbf{25}, 84
(1970)


\bibitem{Ou90a}
Z. Y. Ou, L. J. Wang, L. Mandel, 
Phys. Rev. A, \textbf{40}, 1428 (1989)
\bibitem{Ou90b}
Z. Y. Ou, L. J. Wang, X. Y. Zou, L. Mandel, 
Phys. Rev. A, \textbf{41}, 566 (1990)


\bibitem{Men16}
R. Menzel,
arXiv:1609.09621 [quant-ph], (2016)


\bibitem{Ou97}
Z. Y. Ou, 
Phys. Lett. A \textbf{226}, 323 (1997)



\bibitem{Scu82}
M. O. Scully, K. Dr\"uhl, 
Phys. Rev. A \textbf{25}, 2208 (1982)

\bibitem{Scu91}
M. O. Scully, B.-G. Englert, H. Walther, 
Nature \textbf{351}, 111 (1991)


\bibitem{Zou91}
X. Y. Zou, L. J. Wang,  L. Mandel, 
Phys. Rev. Lett. \textbf{67}, 318 (1991)


\bibitem{Lem14}
G. B. Lemos, V. Borish, G. D. Cole, S. Ramelow, R. Lapkiewicz, A. Zeilinger,
Nature, \textbf{512}, 409 (2014)

\bibitem{Kal16}
D. A. Kalashnikov  A. V. Paterova,	S. P. Kulik, L. A. Krivitsky, %
Nature Photonics, \textbf{10}, 98 (2016)

\bibitem{Ata16a}
S. Ataman, 
Eur. Phys. J. D, \textbf{70}, 127 (2016)

\bibitem{Pat16}
A. Paterova, S. Lung, D. Kalashnikov, L. Krivitsky, 
arXiv:1611.01283 [physics.optics] (2016)



\bibitem{Whe78}
J. Wheeler, \emph{Problems in Formulation of Physics}, ed. G. t.
di Francia, (North-Holland, Amsterdam, 1978)

\bibitem{Whe83}
J. Wheeler's ``Law without law'' in \emph{Quantum Theory and
Measurement}, edited by J. A. Wheeler and W. H. Zurek (Princeton
University Press, Princeton, NJ, 1983)

\bibitem{Jac07}
V. Jacques, E. Wu, F. Grosshans, F. Treussart, Ph. Grangier, A. Aspect, J-F Roch, 
Science \textbf{315}, 966 (2007)

\bibitem{Ion11}
R. Ionicioiu, D. R. Terno, 
Phys. Rev. Lett. \textbf{107}, 230406 (2011)


\bibitem{Kai12}
F. Kaiser, T. Coudreau, P. Milman, D. B. Ostrowsky, S. Tanzilli, 
Science, \textbf{338}, 637 (2012)

\bibitem{Ata14b}
S. Ataman, 
Eur. Phys. J. D  \textbf{68}, 317 (2014)

\bibitem{Ata16b}
S. Ataman, 
Eur. Phys. J. D  \textbf{70}, 160 (2016)


\bibitem{Ma15}
X. Ma, X. Yuan, Z. Cao, B. Qi, Zhen Zhang, 
Npj Quantum Information \textbf{2}, 16021 (2016)




\end{thebibliography}
\end{document}